\documentclass[epj]{webofc}
\usepackage[varg]{txfonts}   % Web of Conferences font
\usepackage[rightcaption]{sidecap}
\usepackage{paralist}
\usepackage{color}
\wocname{EPJ Web of Conferences}
\woctitle{CONF12}
\newcommand{\be}{\begin{equation}}
\newcommand{\ee}{\end{equation}}

\begin{document}
\selectlanguage{english}
\title{Constraining gravity with hadron physics:\\
neutron stars, modified gravity and gravitational waves}

\author{Felipe J. Llanes-Estrada\inst{1}
\fnsep\thanks{\email{fllanes@fis.ucm.es}}
%         \and Second author\inst{2} \and Third author\inst{3}
% etc.
}

\institute{Univ. Complutense de Madrid, Dept. F\'{\i}sica Te\'orica I.\\ 
Fac. CC. F\'{\i}sicas, Avda. de las Ciencias 1, 28040 Madrid, Spain.
%\and
%           The second here 
%\and
%           The last address here
}

\abstract{%
The finding of Gravitational Waves (GW) by the aLIGO scientific and VIRGO collaborations opens opportunities to better test and understand strong interactions, both nuclear-hadronic and gravitational. Assuming General Relativity holds, one can constrain hadron physics at a neutron star. But precise knowledge of the Equation of State and transport properties in hadron matter
can also be used to constrain the theory of gravity itself.
 I review a couple of these opportunities in the context of modified $f(R)$ gravity, the maximum mass of neutron stars, and progress in the Equation of State of neutron matter from the chiral effective field theory of QCD.  
}
\maketitle
%%%%%%%%%%%%%%%%%%%%%%%%%%%%%%%%%%%%%%%%%%%%%%%%%%%%%%%%%%%%%%%%%%%%
\section{Introduction} \label{intro}
%%%%%%%%%%%%%%%%%%%%%%%%%%%%%%%%%%%%%%%%%%%%%%%%%%%%%%%%%%%%%%%%%%%%
\subsection{A convincing discovery}
%%%%%%%%%%%%%%%%%%%%%%%%%%%%%%%%%%%%%%%%%%%%%%%%%%%%%%%%%%%%%%%%%%%%

The famous aLIGO signal~\cite{Abbott:2016blz} convincingly compared with Numerical Relativity predictions, and both LIGO detectors reported it, with matching (inverted) relative phase and a consistent delay. A second event~\cite{Abbott:2016nmj} has confirmed the finding.
The first one,  GW150914, is believed to be caused by a black hole-black hole (BH-BH) merger, with masses $\sim 36M_\odot$ and $29 M_\odot$ yielding a joint BH of $\sim 62M_\odot$ and about $\sim 3M_\odot$ of radiated gravitational energy. The second, GW151226, appears to be the merger of two objects of masses $\sim 7$ and $\sim 14$ $M_\odot$ respectively. All of them are believed to be (quasi) black holes because the competing compact objects available, neutron stars (NS), 
cannot be so heavy  in General Relativity (GR) with conventional understanding of the Equation of State (EOS)  (see subsec.~\ref{subsec:nstarmass}).

Binary neutron star NS-NS mergers are likely to be found within the first three aLIGO runs~\cite{Baiotti:2016qnr}. For the time being, none has been seen out to a distance of 70 Megaparsec (100 MPc in the case of NS-BH merging). Now, our galaxy should contain $O(10^8-10^9)$ pulsars, of which 2300 are already known; a dozen binary ones have orbital periods of order hours.
Based on these estimates of population and the galaxy number density (0.055/MPc$^3$), various studies have predicted between 0.2 and 200 NS-NS merger detections per year.
The merger rate can also be predicted by noticing that Europium is produced by the r-process in amounts of $O(10^{-5}M_\odot)$ per merger~\cite{Vangioni:2015ofa}, yielding an estimated detection rate of 2.5-11 yr$^{-1}$.

Should such a neutron star collision be identified, the most interesting parts of the GW signal at aLIGO would be the later times when the signal tracks the damped oscillations of the merged system, and just before but near the touch down, when tidal disruption of the stars can be large. These stages will be much more sensitive to the hadron matter of the star than the earlier radiation caused by orbital decay.

In the two published events, that final oscillation can be described in terms of 
GR applied to a distorted BH~\cite{Khanna:2016yow}, but exotic physics beyond GR may be used to fake the waveform~\cite{Cardoso:2016oxy}. Thus, the accumulation of events by the LIGO collaboration and other GW detectors can be used to test General Relativity and constrain the parameters of models that try to extend it (section~\ref{sec:constrainG}). As long as one deals with BH-BH collisions, the tests of GR do not require feedback from the Quantum Chromodynamics (QCD) community. But once NS-BH or NS-NS mergers are detected, this will change.

%%%%%%%%%%%%%%%%%%%%%%%%%%%%%%%%%%%%%%%%%%%%%%%%%%%%%%%%%%%%%%%%%%%%%%%%%
\subsection{What is in the signal?}
%%%%%%%%%%%%%%%%%%%%%%%%%%%%%%%%%%%%%%%%%%%%%%%%%%%%%%%%%%%%%%%%%%%%%%%%%

\begin{figure}[h]
\includegraphics[width=\textwidth]{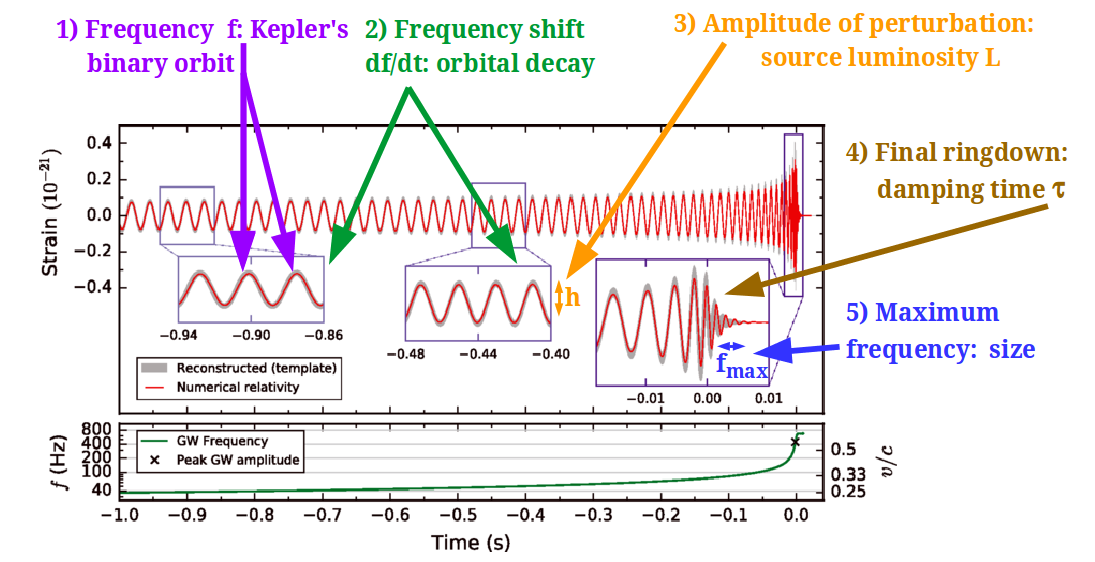}
\caption{\label{fig:GWpulseExplained} Gravitational Wave pulse from GW151226.
Adapted from~\cite{Abbott:2016nmj} under Creative Commons 3.0 license.}
\end{figure}

\begin{SCfigure}[0.9][h]
\includegraphics[width=0.5\textwidth]{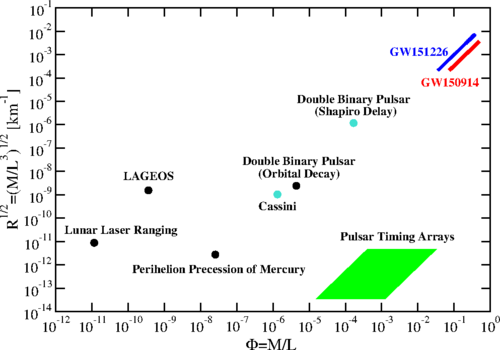}\hspace{0.2cm}
\caption{\label{fig:curvatureregime} The two Gravitational Wave events detected by the aLIGO collaboration probe the largest curvature regime to date, obviously outcompeting all solar system tests both in gravitational potential intensity and spacetime curvature (which vanishes in the Schwarzschild metric outside a star in GR, so would-be dimensional estimates are given instead in case new physics was there). But they are also more extreme than the standing record of the two-solar mass neutron star found by the method of Shapiro delay in a binary system, recalled below in subsection~\ref{subsec:nstarmass}.
 Reproduced from~\cite{Yunes:2016jcc} with permission. \vspace{0.2cm}}
\end{SCfigure}

Figure~\ref{fig:GWpulseExplained} shows the most important pieces of physical information about the collapsing binary system that are contained in a GW pulse, here the second one detected by aLIGO, GW151226.
Circulating clockwise from the top-left corner, these are
\begin{compactenum}
\item The signal frequency with which aLIGO's interferometer arms oscillate: it equals twice the orbital frequency, $\color{blue}f\color{black}=2f_{\rm Kepler}$ and is a piece of data for the two-body orbital problem.
\item The drift of the frequency $\color{blue} \frac{df}{dt}\color{black}$ over many periods 
(which was key to the indirect proof of GW existence in the Hulse-Taylor PSR B1913+16 binary  pulsar): from GW emission theory, this can be used to reconstruct the binary's ``chirp'' mass, 
\be
{\mathcal M} = \left( \frac{(m_1m_2)^3}{m_1+m_2} \right)^{1/5}
            = \frac{c^3}{G} \left(\frac{5\pi^{-8/3}}{96} \color{blue} f^{11/3} \frac{df}{dt} 
\color{black} \right)^{3/5}  \ .
\ee
The chirp mass provides an absolute mass normalization (lacking in the purely Keplerian reconstruction) that allows to state the masses of the binary companions.
\item The amplitude of the perturbation, $\color{blue}h\color{black}\sim 10^{-21}$, gives the distance to the GW source if its luminosity is known: the luminosity $L$ is obtained from the orbital decay. Thus, we can tell  how far the source is from Times Square: $440\pm190$ Megaparsec. 
\item The final ringdown in GW150914 had an attenuation time of $\color{blue}\tau\color{black} \simeq 4$ milliseconds (compare with the Schwarzschild-radius light-crossing time of $2\times 370 $km$/c\simeq 2.5$ ms). 
In the case of a BH-BH merger, the damping is due to the emission of gravitational waves; but in uncollapsed NS mergers yet to be discovered,  viscous damping might have an effect, so that $\tau^{-1}=\tau^{-1}_{GW}+\tau^{-1}_{\eta}$. For a Newtonian sphere of fluid~\cite{Yunes:2016jcc},
$\eta \sim 5\rho R^2 /\tau_\eta$. If we ignored $\tau^{-1}_{GW}$, $\eta$ would be about 4$\times 10^{28}$ Poise for GW150914, far above the $\sim 10^{15}$ Poise expected of neutron matter at T=10 MeV but commensurate with the Alfven viscosity in a huge magnetic field $B\sim 10^{16}$ Gauss. Interesting hadron physics of transport coefficients~\cite{Tolos:2014wla} may hide in NS-NS mergers.   
\item Once the two objects have merged and the final ringdown occurs, the maximum signal frequency $\color{blue} f_{\rm max} \color{black}$ reads off the size of the resulting compound (how fast can a relativistic football spin? $2\pi R_{\rm NS}f_{\rm max}\sim c$; for $R\sim10-100$ km, $f_{\rm  max}$ is in the audible kHz frequency range).
\end{compactenum}
All these properties of the compact objects can be studied with hadron physics feedback once Neutron star mergers are identified. For the time being, pure GR tests are being conducted in the most extreme regime yet probed (see figure~\ref{fig:curvatureregime}).

%%%%%%%%%%%%%%%%%%%%%%%%%%%%%%%%%%%%%%%%%%%%%%%%%%%%%%%%%%%%%%%%%%%%
\subsection{Simultaneous electromagnetic detection}
%%%%%%%%%%%%%%%%%%%%%%%%%%%%%%%%%%%%%%%%%%%%%%%%%%%%%%%%%%%%%%%%%%%%

The simultaneous detection of an electromagnetic signal and gravitational waves from a merger would be very exciting and simultaneous searches are envisioned.
No optical counterpart to GW150914 has been found~\cite{Smartt:2016oeu}. 
An early claim of a simultaneous $\gamma$-ray burst detection by Fermi-GBM has been challenged
because the best positioned detectors saw nothing~\cite{Xiong:2016ssy}, because of sufficiently intense background~\cite{Greiner:2016dsk}, and because Integral/ACS failed to report a sighting. 

Radio astronomers have also failed to detect the GW sources GW150914 and GW151226
(which are expected to be too feeble sources of radio waves)~\cite{Palliyaguru:2016kgg}.

On the hopeful side, a recent study~\cite{Sun:2016pcb} suggests that X-ray emissions from ($\gamma$-quiet) GW-source mergers will be detected by future X-ray observatories such as \emph{Einstein probe}, in the tens of events per year and steradian. One cannot overstate how important it would be to have this additional window to observe the mergers, together with GW or standing alone.

%%%%%%%%%%%%%%%%%%%%%%%%%%%%%%%%%%%%%%%%%%%%%%%%%%%%%%%%%%%%%%%%%%%%
\section{Using Grav. waves and N-star parameters to constrain neutron matter}
%%%%%%%%%%%%%%%%%%%%%%%%%%%%%%%%%%%%%%%%%%%%%%%%%%%%%%%%%%%%%%%%%%%%
\subsection{Constraining the Equation of State with Gravitational Waves}
%%%%%%%%%%%%%%%%%%%%%%%%%%%%%%%%%%%%%%%%%%%%%%%%%%%%%%%%%%%%%%%%%%%%
GW detection at a retarded time $t$ ($t'=t-|{\bf x}-{\bf y}|/c$) provides access to the source's momentum-stress tensor at $t'$, in linear approximation
\be
\bar{h}_{\alpha\beta}(t,{\bf x}) = \frac{4G}{c^4} \int d^3 {\bf y} 
\frac{{\color{blue} T_{\alpha\beta}(t',{\bf y})}}{|{\bf x}-{\bf y}|}\ .
\ee
Of course, ${\color{blue} T_{\alpha\beta}(t',{\bf y})}$ is an object of study in hadron physics.
Most of the literature concentrates on an isotropic ideal fluid with $T^{00}=\rho$, $T^{ij}=P\delta^{ij}$ and $T^{0i}=0$, with density and pressure linked by the EOS $P(\rho)$.
But hadron physics can be much more complex, possibly inhomogeneous~\cite{Buballa:2015awa} or anisotropic~\cite{LlanesEstrada:2011jd}, and dissipative~\cite{Tolos:2014wla}, so that
\be
T = \underbrace{\ \ \ T^{(0)}\ \ \ }_{{\rm Perfect,\, isotropic\, EOS\,} P(\rho)} + \tau^{\rm dissipative} + T^{\rm anisotropic}\ .
\ee

\begin{SCfigure}[0.9]
\includegraphics[width=0.5\textwidth]{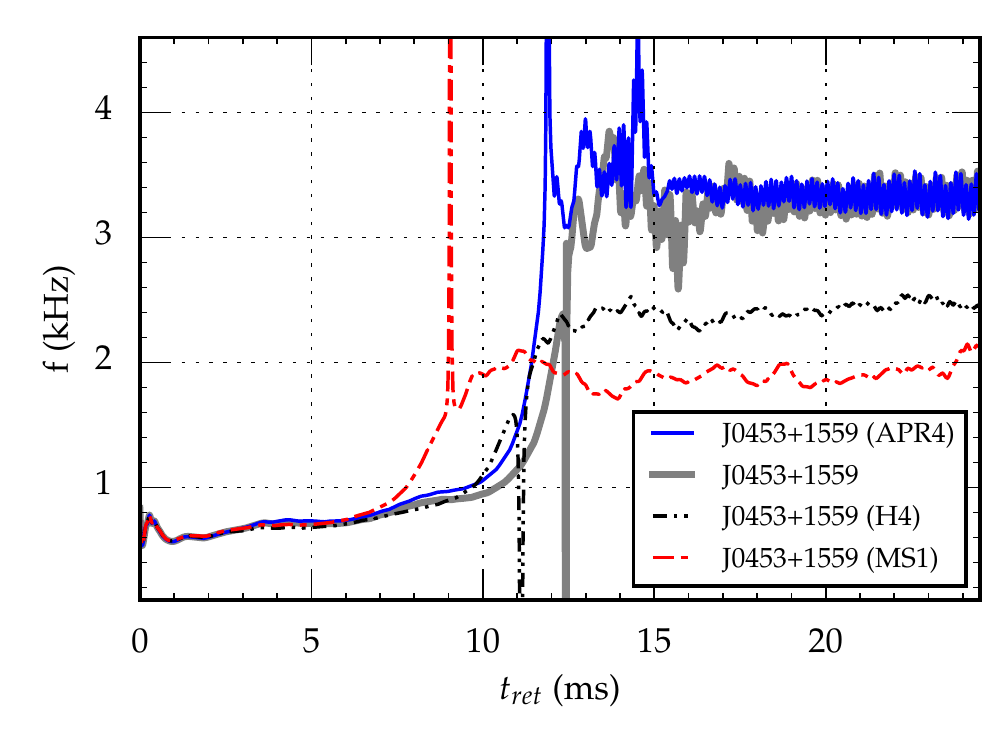}
\hspace{0.05\textwidth}
\caption{\label{fig:dependenceGWonEOS} Three models of the EOS are used to predict the instantaneous frequency (understood as the Fourier transform of the signal over small time intervals) of GW pulses from double neutron-star mergers~\cite{Feo:2016cbs}. 
Discovery of such NS-NS mergers would therefore impose taxing constraints on model computations of the neutron matter EOS. Other simulations~\cite{Chatziioannou:2015uea} suggest that aLIGO will discover/rule out pure quark stars and possibly help discriminate other situations (hybrid or hyperon-containing stars, various condensates, etc.).
Reprinted from~\cite{Feo:2016cbs} with permission.\vspace{0.2cm}}
\end{SCfigure}

In any case, even focusing on the EOS alone, it surprises the hadron physicist that the gold-plated standard in astrophysical applications is the 1998 EOS of~\cite{Akmal:1998cf} based on the Argonne NN potential supplemented by a 3-body force. Other frequently used EOS are the Skyrme-Lyon interaction~\cite{Douchin:2001sv}, various potential-model EOS with hyperons~\cite{Lackey:2005tk}, or relativistic mean field models~\cite{Muller:1995ji}. Astrophysics practitioners do not have the experience to follow the nuances nor the more modern approaches based on Chiral Effective Field Theories (EFT), so they often resort to the free Fermi gas for neutron matter. More sophisticated simulations swipe several of these EOS:
figure~\ref{fig:dependenceGWonEOS}, from~\cite{Feo:2016cbs}, shows how different the ``instantaneous'' frequency  of the emitted GWs is depending on the EOS. 
Clearly, if QCD theory is unable to provide an accurate EOS, the finding of an NS-NS merger will help in constraining it \emph{a posteriori}. Quoting off the Parma-Louisiana collaboration~\cite{Maione:2016zqz},
``\emph{The true EOS for nuclear matter in an environment similar to a
neutron star is not known. (...) We thus have to simulate the
effect of different plausible EOSs on observable quantities, in the
hope to learn about the EOS indirectly through observations.}''
It behooves hadron theory to improve on this situation.

%%%%%%%%%%%%%%%%%%%%%%%%%%%%%%%%%%%%%%%%%%%%%%%%%%%%%%%%%%%%%%%%%%%%
\subsection{Tidal deformability}\label{subsec:tidal}
%%%%%%%%%%%%%%%%%%%%%%%%%%%%%%%%%%%%%%%%%%%%%%%%%%%%%%%%%%%%%%%%%%%%

The luminosity (infered from data as explained in figure~\ref{fig:GWpulseExplained}), for weak radiation, satisfies Einstein's second quadrupole formula, 
\be \label{luminosity}
L = 5 \frac{G}{c^5} \color{blue} \langle \dddot Q_{ij} \dddot Q_{ij}\rangle \color{black}
\ee
that the reader should compare with the usual dipole radiation in electrodynamics,
\be
L_{EM} = \frac{2}{3} \frac{K}{c^3} \langle \ddot D_i \ddot D_i \rangle\ .
\ee
Therefore, gravitational wave pulses just before a merger carry information about the matter
at the source through the quadrupole $Q$.

\begin{SCfigure}[0.9]
\caption{\label{fig:tidal}
Uncertainty in the extraction of the tidal deformability $\Lambda$ from discriminating two Numerical Relativity simulations of presumed to-be data. Though the signal is dominated by the quadrupole due to separation of the two companion stars, careful comparison of the simulated waveforms allows extraction of $\Lambda$ for a given star as shown (here, for $M_{\rm NS} = 1.35M_\odot$), from the  two template waveforms $h_i$. Figure from~\cite{Hotokezaka:2016bzh} reprinted with permission. \vspace{0.5cm}
}
\includegraphics[width=0.5\textwidth]{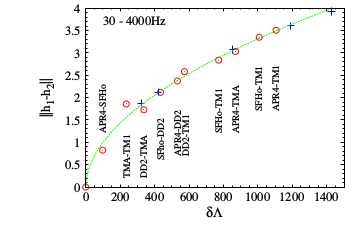}
\end{SCfigure}

Study of the quadrupole in Eq.~(\ref{luminosity}) suggests that one can extract the \emph{tidal deformability} $\Lambda$ (see figure~\ref{fig:tidal}), defined as the coefficient of proportionality, in linear response, of the induced quadrupole of a neutron star to the induced tidal stress due to its binary companion,
\be
Q_{ij}= {\color{blue} \Lambda} E_{ij} \ .
\ee
up to an error $\delta \Lambda$ that depends on the uncertainty $\delta R$ on the neutron star radius~\cite{Hotokezaka:2016bzh}, 
\be
\frac{\delta R /0.91{\rm km}}{R/13{\rm km}}  =  \frac{\delta \Lambda/400}{\Lambda/1000} \ .
\ee
If the tidal deformability of a neutron star becomes known, it will be a challenge to theorists  to calculate it from first principles and will serve as one more constraint on the neutron matter therein.

Taking as reference an aLIGO detection rate of some 10 NS-BH mergers per year~\cite{Abadie:2010cf},  the tidal deformability  $\Lambda$ may be constrained in order of magnitude from a single observed NS-BH merger~\cite{Kumar:2016zlj} or to $O(10\%)$ from 25-50 observations combined, by studying  the ratio of  gravitational wave signals $h_{NS-BH}/h_{BH-BH}$, whose magnitude and phase can be simulated.

%%%%%%%%%%%%%%%%%%%%%%%%%%%%%%%%%%%%%%%%%%%%%%%%%%%%%%%%%%%%%%%%%%%%
\subsection{Neutron star mass and radius} \label{subsec:nstarmass}
%%%%%%%%%%%%%%%%%%%%%%%%%%%%%%%%%%%%%%%%%%%%%%%%%%%%%%%%%%%%%%%%%%%%

Neutron stars in hydrostatic equilibrium need the pressure to compensate the weight of the upper layers as dictated by the Tolman-Oppenheimer-Volkoff equation for a static, spherical body:
\be \label{TOV}\textcolor{blue}{\frac{dP}{dr}} = \textcolor{blue}{- \frac{G_N}{r^2}}
\frac{(\textcolor{blue}{\varepsilon(r)}+P(r))(\textcolor{blue}{M(r)}+4\pi r^3P(r))}{1-\frac{2G_NM(r)}{r}} \ .
\ee
This is supplemented with the EOS $P=P(\rho)$ and the fact (true in GR, but see figure~\ref{fig:massinfR} for generalizations thereof) that the \emph{quantity of matter} in the star coincides with the Schwarzschild mass, 
$M_{GR} =\int_{0}^{R_{\odot}}\,4\pi r^2 \epsilon(r) {\rm d}r$.
Eq.~(\ref{TOV}) is reasonably easy to understand upon comparing with the Newtonian hydrostatic equilibrium equation (the blue color online helps identifying it, by taking $\varepsilon\to M$, $P<<\rho$). Numerical solution of Eq.~(\ref{TOV}) up to an $R_{\rm NS}$ star edge defined by $P=0$ provides the star mass and all other static quantities of interest.

Now, causality sets a limit on the achievable pressure (see later on the caption of fig.~\ref{fig:EOS1}), but nothing limits the amount of matter that may fall on the neutron star. So the pressure picks up
until  the compression is so large that $R_{\rm NS}< 2M_{\rm NS}$ (the Schwarzschild radius) at which point gravitational collapse ensues. Thus, \emph{within General Relativity}, neutron stars have a maximum possible mass -- one can evade the bound~\cite{Astashenok:2014nua} if gravity is modified, as we will discuss later in subsection~\ref{subsec:nstarmass2}. To know it accurately requires also accurate knowledge of the EOS. In early work with a free Fermi gas this $M_{\rm max}$ was $\sim 0.7M_\odot$; it was pinned below about 2$M_\odot$ for long, with most of the pulsar population clustering around $1.4M_\odot$, and is now believed to be above $2M_\odot$ but well below $3M_\odot$. In fact, two-solar mass neutron stars have been reported
in binary systems, by the Shapiro light-delay method~\cite{Demorest:2010bx} and from binary orbital data~\cite{Antoniadis:2013pzd}.

\begin{figure}\centering
\includegraphics[width=0.9\textwidth]{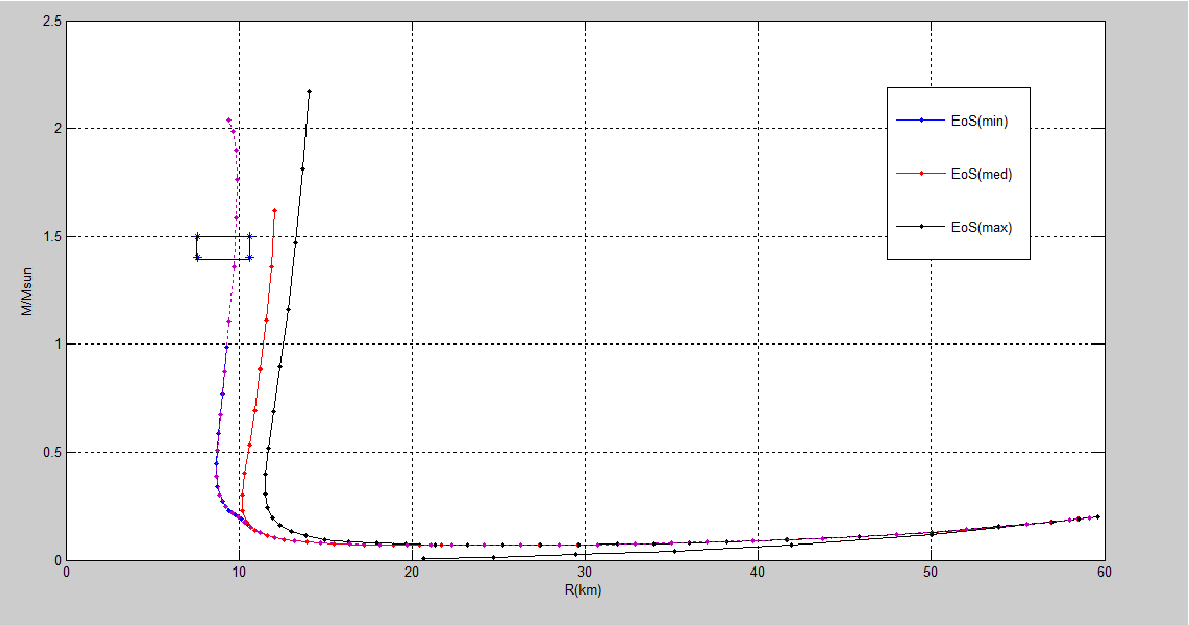}
\caption{\label{fig:vertical}  For each theoretical EOS provided, here the three of~\cite{Hebeler:2013nza}, there is a family of possible stars that trace a curve in the $(R_{\rm NS},M_{\rm NS})$ plane. It is rather vertical near physical pulsar masses, so that good measurements of the star radius are very constraining of the EOS (within the GR framework).
Figure courtesy of Miguel Aparicio Resco; presented as part of his MSc thesis to the faculty of Univ. Complutense de Madrid.}
\end{figure}

As for the stellar radius, quiescent X-ray sources~\cite{Guillot:2014lla} have produced 
$R_{\rm NS} = 9.4\pm 1.2$ km, a number that is uncomfortably low for standard Neutron Star theory (rather predicting 11-13 km; see fig.~\ref{fig:vertical} where the small rectangular box corresponds to this measurement).
The generic reasoning starts with the X-ray luminosity apparent from Earth,
$\color{blue} L_{\rm apparent} \color{black} d^2 = L_{\rm EM} R_{\rm NS}^2$. The right side of this equation contains two unknowns, the size (that we seek) and the actual star luminosity. The distance $d$ is assumed to be known from other measurements (what galaxy hosts the X-ray source).
To extract $R_{\rm NS}$, one needs an additional relation. This is obtained by a thermal fit of the X-ray spectrum. With the extracted temperature one can use Stefan's law for the absolute luminosity, $L_{\rm EM} = \sigma (\pi R_{\rm NS}^2) T^4$ and $R_{\rm NS}$ can then be extracted.

The disagreement with theory has prompted systematic studies that blame absorption and reradiation by the stellar atmosphere, and poor fits to a thermal spectrum; a competing method is using thermonuclear-burst sources as opposed to quiescent stars, and yields $R_{\rm NS}\in (10.4,12.9)$ km, more in line with theory expectations~\cite{Steiner:2012xt}.  Here, GW measurements of the tidal deformability (discussed in subsec.~\ref{subsec:tidal}) or the maximum spinning frequency of a light enough merger endproduct (figure~\ref{fig:GWpulseExplained}) could bring a new measurement with totally different systematics. 

Together, mass and radius measurements of the same neutron star or separate measurements over a significant population are already very constraining of the EOS. This is because the mass-radius diagram curve that comes from solving Eq.~(\ref{TOV}) for each EOS displays very vertical slopes (practically constant radii) in the region of interest, as shown in figure~\ref{fig:vertical}.

Finally, I have not discussed the spin of pulsars nor their glitches, but its impact on the star's core (the realm of hadronic physics) is second to that on the (nuclear) physics of the crust and its entrainment. Fascinating topics such as superfluid vortices, starquakes and more are used to study the topic.  Also the cooling and other nonequilibrium processes in neutron stars have produced much discussion. My ignorance here is vast.

%%%%%%%%%%%%%%%%%%%%%%%%%%%%%%%%%%%%%%%%%%%%%%%%%%%%%%%%%%%%%%%%%%%%
\section{Constraining gravity with neutron star data and hadron theory}\label{sec:constrainG}
%%%%%%%%%%%%%%%%%%%%%%%%%%%%%%%%%%%%%%%%%%%%%%%%%%%%%%%%%%%%%%%%%%%%

%%%%%%%%%%%%%%%%%%%%%%%%%%%%%%%%%%%%%%%%%%%%%%%%%%%%%%%%%%%%%%%%%%%%
\subsection{Constraints \emph{without} hadron theory}
%%%%%%%%%%%%%%%%%%%%%%%%%%%%%%%%%%%%%%%%%%%%%%%%%%%%%%%%%%%%%%%%%%%%
So far I have assumed that General Relativity is the correct theory of gravity and that data can, within its framework, inform hadron physics. Now I take the opposite stance: 
dark energy, dark matter, inflation, the lack of a consistent and empirically sensible quantization of gravity, or the prediction of spacetime singularities, are varyingly unsatisfactory features of GR and numerous attempts have been made at generalizing it 
(massive gravity, scalar-tensor and $f(R)$ theories, Chern-Simons, Horava-Lifschitz and many others, see~\cite{Joyce:2016vqv}). 
Here, orbital measurements are of great use; for example, those in the J0348+0432 2$M_\odot$ pulsar with a white dwarf companion~\cite{Antoniadis:2013pzd} constrain the scalar-tensor coupling ratio of Brans-Dicke theories from the measured orbital decay frequency shift $df/dt$.

\begin{SCfigure}[0.9]
\includegraphics*[width=0.5\textwidth]{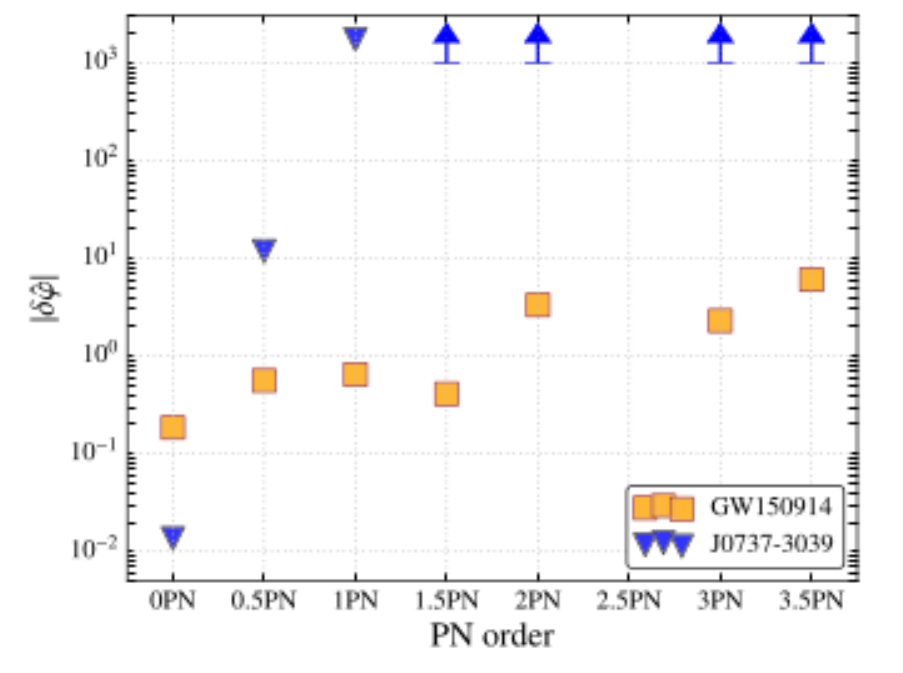}
\hspace{0.05\textwidth}
\caption{\label{fig:postN}
aLIGO already provides extremely competitive constraints on the postNewtonian parameters. These have the advantage of being practically universal (large classes of theories can be cast in the postNewtonian framework for small separations from flat spacetime). The disadvantage of the analysis is that it cannot probe large strides from General Relativity, as it is perturbatively close to it; but constructing a very different theory that gives similar results to GR is, at least, contrived. 
Reprinted from~\cite{TheLIGOScientific:2016src} with permission.
\vspace{0.5cm}}
\end{SCfigure}

If spacetime is not extremely curved, the post-Newtonian approximation (a classical EFT)
is very often used; in fact, aLIGO data are already being put to use (see figure~\ref{fig:postN})
to constrain the postNewtonian parameters and are having spectacular impact.
The new constraints are a factor $O(10^3)$ tighter with the GWs from BH-BH merger
than with earlier binary-pulsar orbital period measurements (these probe the Schwarzschild metric outside either star).

Now, generically thinking, the community has been constraining hadron EOS from neutron star data assuming the validity of General Relativity. The EOS is known in conventional nuclei, and the extrapolation (in density) needed for neutron star interiors is a factor of 2-5. However, the gravitational acceleration outside the star (where binary measurements have constrained GR directly), is $g\sim 300$m/s$^2$; in the interior of white dwarves, $g\sim 10^6$m/s$^2$; and inside a neutron star, $g\sim 10^{12}$m/s$^2$. Going from the former to the latter
requires extrapolating gravity over 10 and 6 orders of magnitude, respectively; it seems more sensible to use computations in reverse gear, and put to use everything that is known about hadron physics (a much smaller extrapolation) for constraining gravity.
The dichotomy can be seen immediately from the strong Equivalence Principle and Einstein's equations,
\be
G_{\mu\nu} + \lambda g_{\mu\nu} = \frac{8\pi G}{c^4} T_{\mu\nu}\ ;
\ee
are any putative disagreements of theory and experiment to be assigned to the left side (gravity) or to the right side (hadrons)? The task of hadron theory in this field is to make sure that $T$ is well understood, to explore the left side.

In the few remaining pages, where hadron physics comes back into play, I shall limit myself to extensions of General Relativity that modify its action but respect all its symmetries and basic spacetime setup; giving up Lorentz or discrete symmetries requires separate discussion~\cite{Tasson:2016xib}.

%%%%%%%%%%%%%%%%%%%%%%%%%%%%%%%%%%%%%%%%%%%%%%%%%%%%%%%%%%%%%%%%%%%%
\subsection{Maximum mass of a neutron star and energy available for GW emission} \label{subsec:nstarmass2}
%%%%%%%%%%%%%%%%%%%%%%%%%%%%%%%%%%%%%%%%%%%%%%%%%%%%%%%%%%%%%%%%%%%%

\begin{SCfigure}[0.9]
\includegraphics*[width=0.5\textwidth]{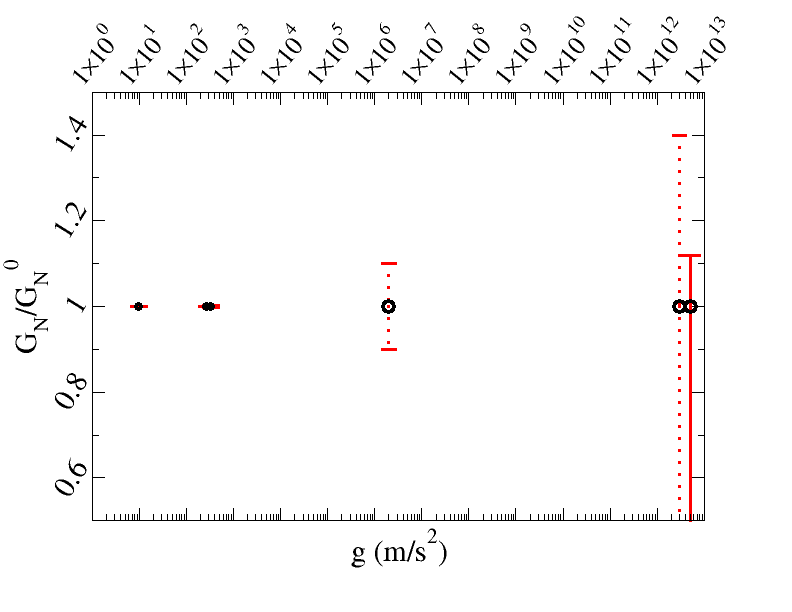}
\hspace{0.05\textwidth}
\caption{\label{fig:varyG}
Constraints on variations of Cavendish's constant $G$, without modifying GR. Our computation is the rightmost one at the highest gravitational acceleration, and is obtained directly by solving Eq.~(\ref{TOV}) with the EOS of figure~\ref{fig:EOS1}. Weakening $G$ yields arbitrarily large $M_{\rm NS}$; but increasing $G$   makes  a $2M_\odot$ neutron star unreachable~\cite{Dobado:2011gd}. 
The more conservative error bar assumes that the EOS is only reliable to a smaller Fermi neutron momentum $k_F\sim 400$ MeV, the smaller error bar was computed with $k_F\sim 600$ MeV.
The reference $G$ is taken on Earth; other constraints come from solar system tests, binary pulsars and white dwarf stars. 
}
\end{SCfigure}

The first exercise that I mention, based in~\cite{Dobado:2011gd}, consists in leveraging the maximum neutron star mass discussed in subsection~\ref{subsec:nstarmass}: since 2$M_\odot$ stars have already been detected, a strengthening of gravity (that forces earlier star collapse and thus reduces the maximum mass) is limited. This results in a limit on the value of the Newton-Cavendish constant $G$ inside the star as can be seen in figure~\ref{fig:varyG}.  

A different situation arises if one allows for modifications of General Relativity.
In~\cite{Resco:2016upv}, we have explored neutron stars within $f(R)$ theories. These are very popular to account for dark energy in cosmology: this is the limit of very tenuous density but very large sizes~\cite{delaCruz-Dombriz:2016bqh}. Neutron stars are rather on the opposite end of GR, compact objects with intense fields. Still $f(R)$ theories can be used to model inflation, and neutron stars are a step in the ladder of phenomena at larger $R$ that leads to it, so they can be used to put constraints on them~\cite{Capozziello:2011nr}. 
They are introduced as very simple modifications of the Einstein-Hilbert action,
\be
S=\frac{1}{16\pi G}\int {\rm d}^4 x \sqrt{-g}\,[R+\color{blue} f(R)\color{black}]
\ee
but the resulting Einstein's differential equations are of higher order and technically very challenging.

From several viable ones~\cite{Bamba:2010zz} let me focus here on $R+f(R)=R+aR^2$ because, surprisingly, the ``Starobinsky'' inflation that it describes is favored by the Planck collaboration data. This is because it predicts a smaller tensor/scalar ratio in cosmic microwave background perturbations than other alternative models. 

\begin{figure}
 \includegraphics[width=0.3275\textwidth]{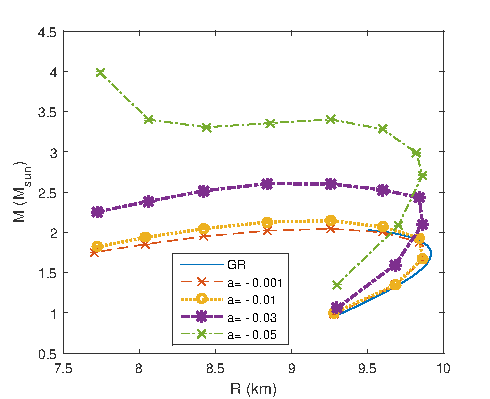} 
 \includegraphics[width=0.3275\textwidth]{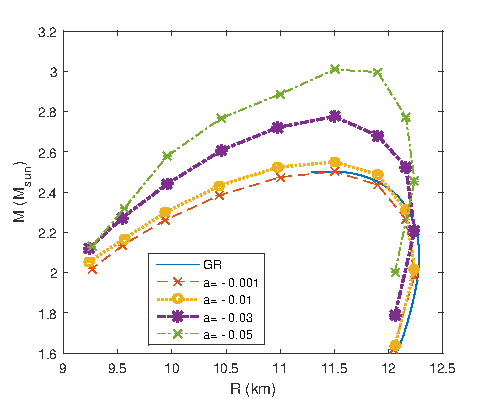}
 \includegraphics[width=0.3275\textwidth]{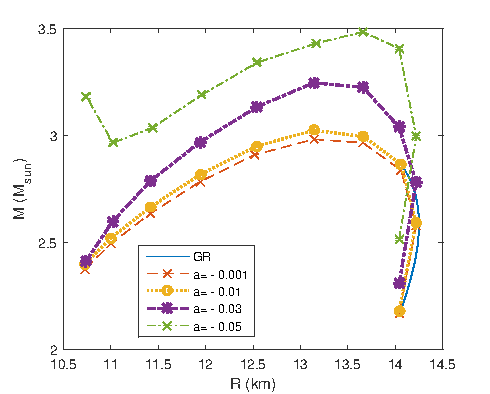}
\caption{\label{fig:MofRmodgrav} The neutron star mass-radius diagram in $f(R)=aR^2$ modified gravity~\cite{Resco:2016upv} with $a=-0.05$.
Neutron stars with masses larger than 2.4$M_\odot$ are easily allowed, so that negative $a$ is unconstrained. Choosing the opposite sign of $a$ however reduces the maximum achievable mass~\cite{Yazadjiev:2014cza}. The EOS employed are those of~\cite{Hebeler:2013nza}, with ``stiffness'' (see subsection~\ref{subsec:EOS2}) growing from the left plot to the right plot.
} 
\end{figure}

Depending on the sign of $a$ that one chooses, one may find solutions with $M_{\rm NS}>3M_{\odot}$, as shown in figure~\ref{fig:MofRmodgrav}.
For negative $a$ it is easy to find solutions of larger mass than allowed in General Relativity, just as in the case of diminishing $G$. For positive $a$ however, one can fail to find solutions reproducing the known $2M_\odot$ pulsars and thus constraints on $f(R)$ are already possible~\footnote{A very interesting discussion is that of superluminal propagation for $a<0$; without violating special relativity, that is built into the formalism, at any time, one can find faster than light propagation of gravity waves, as the dispersion relation for the equivalent scalar mode from Eq.~(\ref{modes})is $m_R^2= \frac{1+f_{'R}(0)}{3 f_{''R}(0)}$. A network of GW detectors beyond aLIGO can constrain this propagation. This apparent tachyons do not appear for $a>0$ but discussing them would take us too far afield.}.

An interesting difference from GR is that solving the equivalent of the Tolman-Oppenheimer-Volkoff system is more convenient with an additional variable, the scalar curvature $R(r)$. We take the metric with the usual static, spherically symmetric form
\be \label{metric}
ds^{2}=B(r)\,dt^{2}-A(r)\,dr^{2}-r^{2}(d\theta^{2}+\sin^{2}{\theta}\,d\phi^2)
\ee
and write down Einstein's equations for $A$, $B$ and the star matter content. The additional equation satisfied by $R$ is
\be
R=\frac{8\pi G\,T-2\,f(R)-3\,\nabla^{\alpha}\nabla_{\alpha}f_{R}}{1-f_{R}}\ ;
\ee
all of them are written together as a first order system, integrated with the Runge-Kutta algorithm from the star center until $P=0$ (star's edge), then continued in a vacuum to a large distance so that the outside solution is at hand and can be matched to the asymptotically flat space.

\begin{SCfigure}[0.9]
\includegraphics*[width=0.5\textwidth]{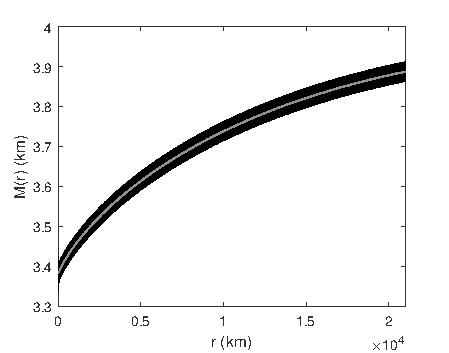}
\hspace{0.05\textwidth}
\caption{\label{fig:massinfR}
If the mass of the stellar system is understood as the one from matching to the Newtonian potential at infinite distance from the star, then it receives contributions from the field outside the star. A way to visualize it is to use the Schwarzschild parametrization of $A(r)$ and $B(r)$ in Eq.~(\ref{metric}) for all $r>R_\odot$ but allow the mass therein to be $r$-dependent, resulting in a function $M(r)$. Note that the star radius is about 10 km, so that the Runge-Kutta numerical integration extends very far from the star until the asymptotic regime is recovered.
\vspace{0.5cm}}
\end{SCfigure}

A second difference to GR is the actual definition of ``mass''. Now, the quantity of matter in the star is not the same number that appears upon  matching the external static metric to Newton's potential at infinite distance: this last apparent gravitational mass receives contributions from vacuum (that can only locally be considered Schwarzschild) out to many star radii, as seen in figure~\ref{fig:massinfR}.

Further, the linearization of metric perturbations to obtain GWs shows that, in addition to the two transverse tensor polarizations, there is a third, scalar mode 
\be \label{modes}
h=\left( 
\begin{array}{cc}
{\color{blue} h_0}+h_{+} & h_{x} \\ 
-h_{x} & {\color{blue} h_0}-h_{+}
\end{array} 
\right)
\ee
so one can see that in certain limits, scalar-tensor theories are equivalent to $f(R)$ metric theories.

A way to think of it is that the static, spherically symmetric solutions in $f(R)$ theories can be tagged with the quantity of matter up to the star's edge, $M_{\rm NS}= M(R_{\rm NS})$. However this tag differs from the tag that we would assign them by Newton's potential at very far distances, $\Phi\to -GM_\infty/r$. The two quantities are equal in GR.

It has not escaped the reader that, if beyond-GR theories allow for larger NS masses than 2-3$M_\odot$, the identification of the GW signals from aLIGO are not necessarily assigned to BH-BH collisions. Likewise, the available energy that can be emitted as gravitational radiation can be very different in $f(R)$ theories and in GR (and part of it in the scalar mode). Much work remains here, though, to attempt to construct realistic waveforms to compare with aLIGO.

%%%%%%%%%%%%%%%%%%%%%%%%%%%%%%%%%%%%%%%%%%%%%%%%%%%%%%%%%%%%%%%%%%%%
\subsection{The Equation of State} \label{subsec:EOS2}
%%%%%%%%%%%%%%%%%%%%%%%%%%%%%%%%%%%%%%%%%%%%%%%%%%%%%%%%%%%%%%%%%%%%
Precise knowledge of the Equation of State from first principles QCD or, as far as possible, the effective theories thereof, is then an endeavor of potential impact in the field of gravity, and this last section is dedicated to inform readers with an astrophysics background on the progress therein, as I view it.

\begin{SCfigure}[0.9]
\includegraphics*[width=0.5\textwidth]{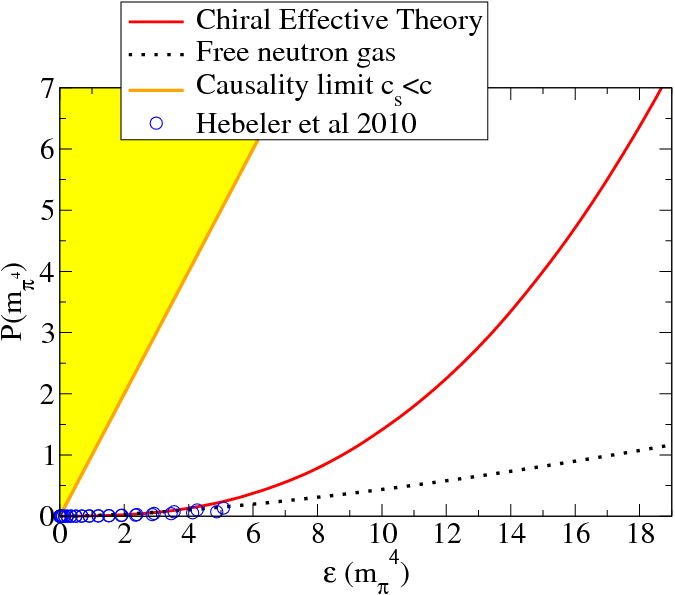}
\hspace{0.05\textwidth}
\caption{Equation of State for neutron matter in chiral EFT+dispersion relations~\cite{Dobado:2011gd} (solid line, red online). The shaded yellow area is immediately excluded by causality (the speed of sound $c_s = \sqrt{\frac{dP}{d\epsilon}}$ exceeds that of light, $c_s\geq 1$). In fact the slope of no EOS in this plot can exceed 1 (that of the line bounding the shaded area). This means that ignorance about the EOS at high density \emph{does not prevent us from constraining gravity} as the EOS will, from any given $\epsilon$ up to which nuclear\& particle physics can be reliably employed, be at most as steep as that line. The Green's function-Monte Carlo EOS of~\cite{Hebeler:2010jx} is very compatible at low energies. Also shown is the free neutron gas (much ``softer'', that is, with lower pressure).\label{fig:EOS1}\vspace{0.2cm}}
\end{SCfigure}

Figure~\ref{fig:EOS1} illustrates the basic physics of the EOS.
The first thing to note, as explained in the caption, is that \emph{the pressure cannot grow arbitrarily fast with density (causality)}. This is the basic feature that allows to put bounds on gravity modifications even with incomplete knowledge. In fact, within GR, just because of causality, $M_{\rm NS}<3M_{\odot}$ approximately. Of course, the constraints will get increasingly better as the EOS is better predicted from theory. 
Shown in the plot, as the dotted line, is the free Fermi gas (which, in spite of much progress in hadron theory, is still used by astrophysicists, such as a recent computation of tidal deformation in neutron stars~\cite{Yu:2016ltf}).

The solid line and circles are two EOS based in modern~\cite{Meissner:2007zza} chiral nuclear forces (EFT). The solid line, from~\cite{Lacour:2009ej}, employs a dispersive analysis to deal with the many-body problem to which the LO chiral Lagrangian is applied, while the circles employ the Green's functions Monte Carlo method.

Notice the units of the graph. MKS or cgs units lead to absurdly large numbers. Very often, MeV/fm$^3$ are employed for the energy density, but this requires two geocentric scales, the electronvolt and the Fermi (and pressure is rarely quoted in MeV/fm$^3$). Since $P$ and $\epsilon$ have equal natural dimensions, I prefer $m_\pi^4$ that only requires one mass scale for both (any other choice of scale would be adequate, but this one makes all magnitudes of order 1 in neutron stars). 

The plot is cut for $\epsilon\sim 20m_\pi^4$ where the red line steepens and breaks causality: obviously the EFT calculation stopped being reliable before then. Now, the reader might believe that the computation based on neutrons should be substituted at high $\epsilon$ of order a few $m_\pi^4$ by more exotic phases: quark phases are very popular~\cite{Fogaca:2016mnw} as are mixed-flavor phases with strangeness (hyperons) that must appear for high enough Fermi levels (as the strange quark allows to relax the Fermi surface since it is one more fermion degree of freedom), or even nonspherical hadrons~\cite{LlanesEstrada:2011jd}. Never mind the reader's preference for exotic QCD physics in a neutron star: the resulting EOS will fall to the right (will be ``softer'') of the neutron-based EFT EOS in that plot. This is precisely because a change of phase relaxes the free energy, and hence lowers the pressure.
So a possible way to obtain bounds on modified gravity is to control very well the equation of state within the neutron-based EFT.

\begin{SCfigure}[0.9]
\includegraphics[width=0.5\textwidth]{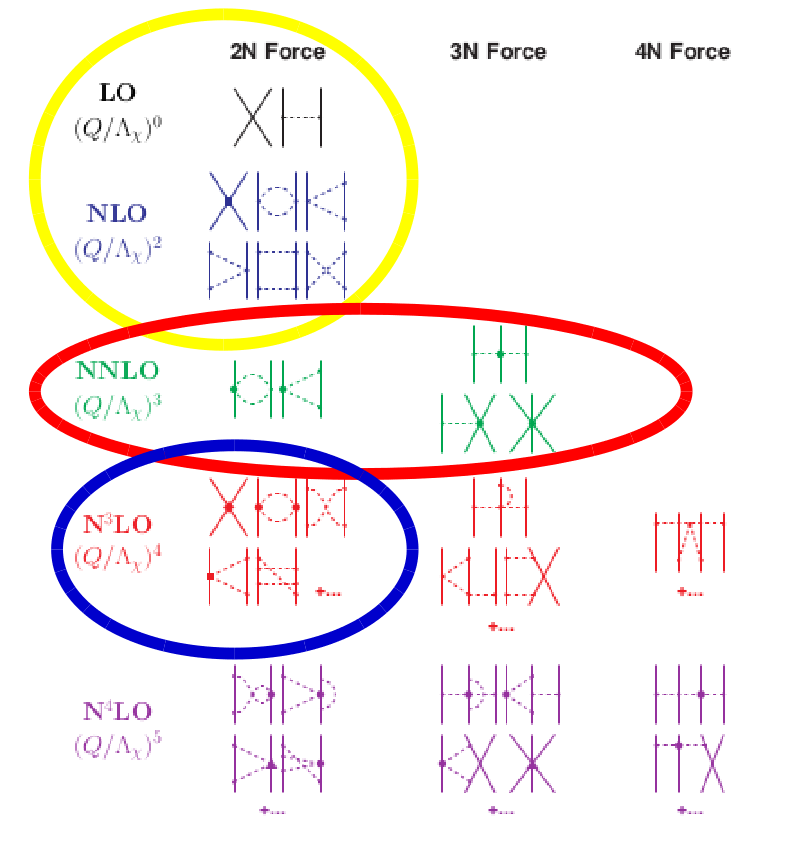}
\caption{The nucleon-nucleon interaction in Effective Theory is generated order by order in the chiral expansion (a low-momentum, small pion-mass expansion). It offers a measure of control over the order at which more complicated potential terms and relativistic, many-body corrections have to be included, and allows for a systematic estimate of their effects and uncertainties. Note particularly that 3-body forces enter only at NNLO. 
Currently, the $NN$ interactions is being worked to sixth order~\cite{Entem:2015xwa} but computations of interest for neutron stars to such high accuracy will have to wait. In fact, a recent criticism~\cite{RuizArriola:2016sbf} raised at this conference is that fitting the chiral EFT parameters to NN scattering \emph{phase shifts} as opposed to \emph{cross section data} is introducing a large error in subsequent nuclear structure calculations that are in poor agreement with experiment. 
Reprinted with permission from~\cite{Sammarruca:2016ajl} under Creative Commons License.\vspace{0.5cm}\label{fig:counting}}
\end{SCfigure}

\begin{SCfigure}[0.9]
\includegraphics[width=0.5\textwidth]{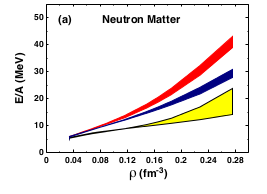}
\caption{\label{fig:EOSEFT}
Although the chiral power counting in neutron matter~\cite{Lacour:2009ej} differs from that in vacuo appropriate for nucleon-nucleon scattering experiments, 
I find a very interesting exercise to obtain the Equation of State of neutron matter at increasing order even with that counting. The shaded lines are colored online according to the order of perturbation theory: the highest interaction terms included in each computation are marked (in the same color) in figure~\ref{fig:counting}. The width of the bands corresponds to varying the cutoff in loop momenta in the interval 450-600 MeV.  
Reprinted with permission from~\cite{Sammarruca:2016ajl} under Creative Commons License.}
\end{SCfigure}

An effort~\cite{Sammarruca:2016ajl} to understand the convergence of the low-momentum  expansion 
of QCD from first principles as applied to the EOS is reported in figures~\ref{fig:counting} and~\ref{fig:EOSEFT}. The EFT approach has several advantages: the terms that appear in the Hamiltonian are known at a given order in the counting, once one has decided what the target precision is. Many-body forces are suppressed and appear at higher orders. The symmetries fix what terms are possible and how many parameters there should be: there are correlations among the long distance parts of the interactions as they stem from a field theory. The short-distance counterterms are not fixed by the EFT but fit to NN scattering and few-nucleon nuclei, but may one day be extracted from the underlying QCD.
 
The inconvenient is that the EFT is renormalizable only order by order and the number of parameters grows eventually exceeding the low energy data and loosing predictive power.
Another issue is that of cutoff dependence~\cite{Arriola:2014fqa}
 (hadron interactions are very strong, near singular at short distances, and upon truncating the expansion not all cutoff dependence is absorbed).

A caveat specific to the computations of~\cite{Sammarruca:2016ajl}, that otherwise show reasonable convergence for moderate densities, is that the chiral counting of the EFT employed is that of hadrons {\it in vacuo}; but because of the additional scale in the neutron star medium, $k_F$, a modified counting is necessary~\cite{Lacour:2009ej}. We look forward to progress in this respect.

NLO and NNLO computations of the EOS are complete.
From N$^3$LO, only the 2-body part has been included; to add the 3-body contributions,
one would first  need to refit the chiral parameters to  triton data, which has not yet been carried out~\cite{Sammarruca:2016ajl,Sammarruca:2014zia}.
A similar exercise in convergence of the perturbative expansion of the EOS has been reported by~\cite{Drischler:2016djf}, with similar results. Also analogous are the computations with the
Functional Renormalization Group computations~\cite{Drews:2016wpi} and the $\chi$EFT.

%%%%%%%%%%%%%%%%%%%%%%%%%%%%%%%%%%%%%%%%%%%%%%%%%%%%%%%%%%%%%%%%%%%%
\section{Conclusion}
%%%%%%%%%%%%%%%%%%%%%%%%%%%%%%%%%%%%%%%%%%%%%%%%%%%%%%%%%%%%%%%%%%%%

I have tried to convince hadron physicists of the merit in pursuing neutron star studies
that may tighten theories and models beyond General Relativity. Particularly, the maximum mass of neutron stars and the recently discovered gravitational waves hold much appeal.
But there are many other observables that can be explored.
Conversely, I hope some programmers of numerical relativity codes find my discussion of more modern Equations of State a starting point to delve into EFT as opposed to dated models, that should only be used when QCD-based approaches are not available or for sanity cross checks. The EFT calculations seem to have gained traction and much more progress is to be expected.

One wonders then, what is of all the exotic phases and additional particles that QCD and quark flavor can bring in? Well, it is obvious that they soften the EOS. Possibly too much, as many have been ruled out by the 2$M_\odot$ stars. But their presence is not optional; (at least for some of them) it must follow (or not) from well defined calculations from the QCD action. 
What I and others guess is that, just as the 3-body force is repulsive, higher terms in the EFT-EOS must also make it stiffer, to later soften into something phenomenologically viable when other degrees of freedom activate. 
Of course, another logical possibility is to have doubly exotic physics: new gravity phenomena (such as $f(R)$ corrections to the Einstein-Hilbert action) providing the margin for the EOS to actually be softer~\cite{Astashenok:2014pua}.

Even if the EFT-EOS is only reliable up to a certain density, there are fundamental restrictions on hadron and hadron matter properties, such as unitarity and causality, and we have used the latter to constrain allowable variations of the gravitation constant $G$ in a neutron star. Likewise, we have studied them in $f(R)$ theories. 
More work is planned to see how much precision can be expected from this line of study in constraining $a$, the parameter of $R^2$ gravity and that enters Starobinsky inflation.

%%%%%%%%%%%%%%%%%%%%%%%%%%%%%%%%%%%%%%%%%%%%%%%%%%%%%%%%%%%%%%%%%%%%
\section*{Acknowledgments}
%%%%%%%%%%%%%%%%%%%%%%%%%%%%%%%%%%%%%%%%%%%%%%%%%%%%%%%%%%%%%%%%%%%%
This presentation was needed to cover such exciting developments at the plenary of the Confinement-XII conference; as convener of its \emph{QCD and New Physics} section E, I substituted for
more qualified speakers, all previously engaged.
My gratitude for their endless work is due to the organizers, and more so to those with which I directly interacted in solving all daily challenges: E. Andronov, Y. Foka, A. Katanaeva, V. Kovalenko, M. Janik; and especially to Nora Brambilla for keeping this forum open over the years. Thanks are due also to my fellow conveners, particularly M. Gersabeck and E. Mereghetti. Work supported by grants UCM:910309 and MINECO:FPA2014-53375-C2-1-P.

%\begin{figure}[h]
%% Use the relevant command for your figure-insertion program
%% to insert the figure file.
%\centering
%\includegraphics[width=7cm,clip]{test.jpg}
%\caption{Please write your figure caption here}
%\label{fig-1}       % Give a unique label
%\end{figure}

%For figure with sidecaption legend use syntax of figure~\ref{fig-2}.
%\begin{figure}[h]
%% Use the relevant command for your figure-insertion program
%% to insert the figure file.
%\centering
%\sidecaption
%\includegraphics[width=7cm,clip]{test.jpg}
%\caption{Adapted from under the Creative Commons Attribution 3.0 License.}
%\label{fig-2}       % Give a unique label
%\end{figure}

%For tables use syntax in table~\ref{tab-1}.
%\begin{table}[h]
%\centering
%\caption{Please write your table caption here}
%\label{tab-1}       % Give a unique label
%% For LaTeX tables you can use
%\begin{tabular}{lll}
%\hline
%first & second & third  \\\hline
%number & number & number \\
%number & number & number \\\hline
%\end{tabular}
%% Or use
%%\vspace*{5cm}  % with the correct table height
%\end{table}

%%%%%%%%%%%%%%%%%%%%%%%%%%%%%%%%%%%%%%%%%%%%%%%%%%%%%%%

\end{document}